\documentclass[journal,transmag]{IEEEtran}
\ifCLASSINFOpdf
\else
\fi
\usepackage{graphicx}
\usepackage{booktabs}

\begin{document}
%
\title{A Comprehensive Analysis of Routing Vulnerabilities and Defense Strategies in IoT Networks}


\author{\IEEEauthorblockN{Kim Jae-Dong\IEEEauthorrefmark{1}}
\IEEEauthorblockA{\IEEEauthorrefmark{1}Republic of Korea Armed Forces under the Ministry of National Defense}
\thanks{This research was supported by a preliminary study funded by a military institute under the Ministry of National Defense, and note that this manuscript will be submitted to National Defense-related journals or conferences in the future. J.-D. Kim is affiliated with Armed Forces agency under the Ministry of National Defense (email: jaedong0731@mnd.go.kr).}}

%



\IEEEtitleabstractindextext{%
\begin{abstract}
The rapid expansion of the Internet of Things (IoT) has revolutionized various domains, offering significant benefits through enhanced interconnectivity and data exchange. However, the security challenges associated with IoT networks have become increasingly prominent owing to their inherent vulnerability. This paper provides an in-depth analysis of the network layer in IoT architectures, highlighting the potential risks posed by routing attacks, such as blackholes, wormholes, sinkholes, Sybil, and selective forwarding attacks. This study explores the unique challenges posed by the constrained resources, heterogeneity, and dynamic topology of IoT networks, which complicate the implementation of robust security measures. Various countermeasures, including trust-based mechanisms, Intrusion Detection Systems (IDS), and routing protocols, are evaluated for their effectiveness in mitigating these threats. This study also emphasizes the importance of considering misbehavior observation, trust management, and lightweight defense strategies in the design of secure IoT networks. These findings contribute to the development of comprehensive defense mechanisms tailored to the specific challenges of IoT environments.
\end{abstract}

}

\maketitle

\IEEEdisplaynontitleabstractindextext

%
\IEEEpeerreviewmaketitle

\section{Introduction}
%
%
%
%

\IEEEPARstart{I}{nternet} of Things (IoT) represents a significant advancement in the digital world, connecting various physical devices and enabling them to communicate and share data. These devices, which range from simple household items to complex industrial machines, are embedded with sensors, software, and other technologies that facilitate the collection and exchange of data over the internet \cite{r1}.  This interconnectivity allows seamless integration and automation across different domains, including smart homes, healthcare, transportation, and industrial automation. The significance of IoT is multifaceted. In smart homes, IoT devices, such as thermostats, lighting systems, and security cameras, can be remotely controlled to enhance convenience and energy efficiency. In healthcare, IoT devices enable the real-time monitoring of patient health, providing critical data for early diagnosis and timely intervention. Industrial applications of the IoT have led to improved production efficiency, predictive maintenance, and better supply chain management. Additionally, IoT's role in smart cities facilitates efficient resource management, reducing traffic congestion, and enhancing public safety. The proliferation of IoT devices has led to an exponential increase in data generation, offering new opportunities for data analytics and artificial intelligence to drive informed decision making \cite{r1, r2}. Fig. 1 presents the total number of IoT devices with and without connectivity from 2010 to 2027 (estimated). As IoT continues to evolve, it promises to revolutionize industries, improve the quality of life, and create new economic opportunities by integrating the physical and digital worlds in unprecedented ways. The Internet of Things (IoT) offers numerous benefits by enabling interconnectivity and data exchange between diverse devices \cite{r2}. However, this rapid expansion also brings forth significant security challenges that must be addressed to ensure the reliability and safety of IoT networks \cite{r3, r4, r5, r6}.

\begin{figure}[tbp] 
\begin{center}
\includegraphics[width=1.0\linewidth]{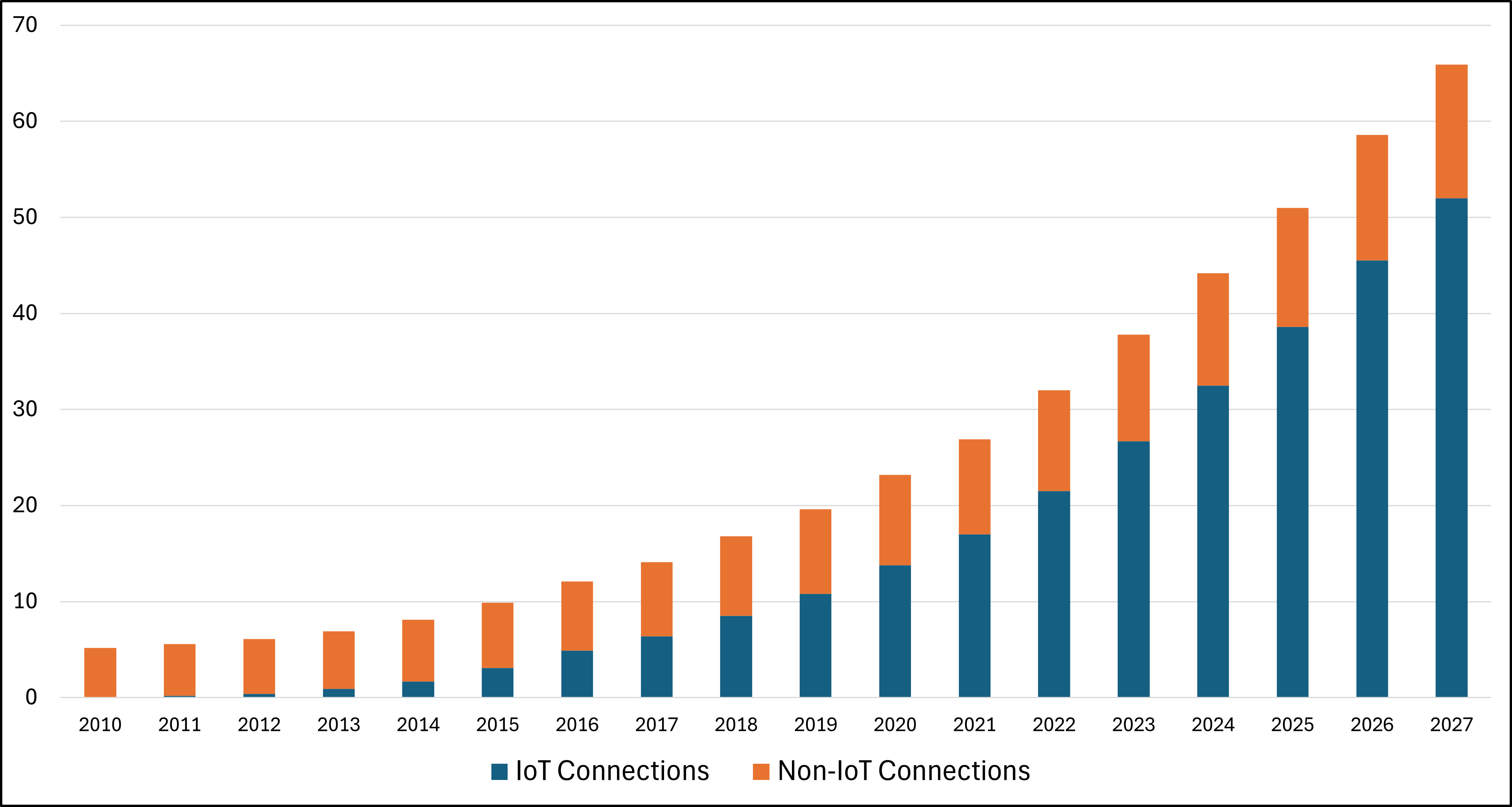}
\end{center}
\caption{IoT Connection versus Non-IoT Connections}
\vspace{-0.6cm}
\label{fig:long}
\label{fig:onecol}
\end{figure}

\begin{table*}[]
\caption{Applications of IoT Networks Across Industries}
\setlength{\tabcolsep}{3pt}
\renewcommand{\arraystretch}{1.3}
\normalsize
\begin{center}
\begin{tabular}{c|c}
\noalign{\smallskip}\noalign{\smallskip}\hline
\hline
\textbf{Domain} & \textbf{Description}                                                                                                                                                                                                \\ \hline
Smart Home      & \begin{tabular}[c]{@{}c@{}}Remote control and automation of home devices, enhancing convenience, energy efficiency, \\ and security.\end{tabular}                                                                  \\ \hline
Healthcare      & \begin{tabular}[c]{@{}c@{}}Real-time health monitoring, data collection, and remote patient management for better \\ healthcare outcomes and personalized treatment.\end{tabular}                                  \\ \hline
Industrial IoT  & \begin{tabular}[c]{@{}c@{}}Improves manufacturing processes with predictive maintenance, real-time monitoring, \\ and efficient supply chain management.\end{tabular}                                              \\ \hline
Smart Cities    & \begin{tabular}[c]{@{}c@{}}Optimizes resource management, reduces traffic congestion, and enhances public safety with \\ smart grids,traffic management systems, and surveillance systems.\end{tabular}           \\ \hline
Agriculture     & \begin{tabular}[c]{@{}c@{}}Enables precision farming, optimizes irrigation and fertilization, and monitors crop health and \\ livestock, increasing yields and resource efficiency.\end{tabular}                   \\ \hline
Transportation  & \begin{tabular}[c]{@{}c@{}}Connected vehicles and IoT-enabled fleet management systems improve safety and efficiency, \\ offering real-time tracking, route optimization, and maintenance scheduling.\end{tabular} \\ \hline
\hline
\end{tabular}
\end{center}
\end{table*}

According to Cisco’s 2024 Cybersecurity Readiness Index, only 3\% of organizations are considered mature in their cybersecurity readiness, with comprehensive security measures to address emerging threats, particularly for IoT devices \cite{r7}. One of the primary security limitations of the IoT is the inherent vulnerability of devices owing to their constrained resources. Many IoT devices are designed with limited computational power, memory, and energy capacity, which restrict the implementation of robust security measures, such as advanced encryption algorithms and real-time security monitoring. This makes IoT devices more susceptible to attacks including unauthorized access, data breaches, and malware infections. Additionally, the heterogeneous nature of IoT networks complicates the establishment of standardized security protocols. IoT devices originate from various manufacturers and operate on different platforms, leading to a lack of uniform security standards. This heterogeneity results in inconsistent security practices and creates potential vulnerabilities across the networks. The dynamic and decentralized nature of IoT further exacerbates security challenges. IoT networks often involve devices that frequently join and leave the network, which makes it difficult to maintain a consistent security posture. This dynamic environment is prone to routing attacks, in which malicious nodes can disrupt communication paths, leading to data interception and network partitioning. Moreover, IoT devices are often deployed in diverse and sometimes hostile environments, which increases the risk of physical tampering and attacks. Unlike traditional computing devices, which are usually secured in controlled settings, IoT devices might be placed in locations with minimal physical security, making them easy targets for attackers seeking to extract sensitive information or sabotage operations \cite{r3, r4, r5, r6}. This study examines various attacks that can occur in the network layer of IoT, including black hole, wormhole, and sinkhole attacks. These attacks disrupt communication paths within IoT networks, leading to data interception and network partitioning. Effective countermeasures such as secure routing protocols and intrusion detection systems are discussed. Additionally, this paper highlights critical considerations when designing these countermeasures, such as resource constraints, network heterogeneity, and dynamic topology changes. Addressing these factors is essential for developing robust and reliable security solutions tailored to the unique challenges faced by IoT networks. This discussion aims to enhance the security and reliability of the IoT systems.

\section{Methodology}
This study employs a structured approach to analyze network layer vulnerabilities in the IoT, focusing on routing attacks and corresponding defense mechanisms. The methodology is divided into three key phases: 1. Literature Review and Threat Identification: A comprehensive review of the existing literature was conducted to identify common routing attacks in IoT networks, such as blackhole, wormhole, Sybil, sinkhole, and selective forwarding attacks. This review provides a foundation for understanding the scope of security challenges and potential defence mechanisms. 2. IoT network simulation and attack scenarios simulating IoT networks were created to model real-world scenarios with diverse devices and dynamic topologies. Various routing attacks have been implemented to assess their impact on network performance, focusing on metrics such as packet delivery, latency, and energy consumption. 3. Defense Mechanism Evaluation Several defense mechanisms, including trust-based systems, Intrusion Detection Systems (IDS), and multipath routing protocols, have been implemented and evaluated. The performance of these mechanisms was measured in terms of detection accuracy, network efficiency, and resource overhead to ensure their suitability for resource-constrained IoT environments. This structured methodology provides a comprehensive framework for analyzing security and performance trade-offs in IoT networks, contributing to the development of effective and scalable defense solutions.

\section{Background Knowledge of IoT Networks}
The Internet of Things (IoT) is a transformative paradigm that reshapes how devices communicate and interact with each other and their environments. At its core, IoT comprises a network of physical objects embedded with sensors, software, and other technologies, enabling them to collect and exchange data. This connectivity facilitates seamless integration across various sectors, improving efficiency, productivity, and user experience. IoT refers to the interconnection of everyday objects via the Internet, ranging from household items to sophisticated industrial equipment. These "smart" devices are designed to gather data from their surroundings, process information, and communicate with other systems. Table I summarizes the fields in which IoT is widely used across various industries. An IoT system is composed of several essential components. First, sensors collect data from the environment, such as temperature, humidity, or motion, while actuators perform actions based on the processed data, such as adjusting a thermostat or turning lights on. Connectivity plays a crucial role, as communication protocols and networks, including Wi-Fi, Bluetooth, ZigBee, and cellular networks, allow these devices to share and exchange data. Once the data are collected, they must be processed and analyzed, either locally on the device (edge computing) or on centralized servers (cloud computing). Finally, a user interface provides the means for users to interact with the IoT system, typically through mobile apps or web dashboards. The IoT offers a broad range of applications, significantly enhancing functionality and efficiency across various domains. These interconnected devices transform traditional processes and enable innovative solutions in diverse fields \cite{r1, r2}.

\subsection{Smart Homes}
The IoT has revolutionized the concept of smart homes, where devices such as thermostats, lighting systems, and security cameras can be controlled remotely via smartphones or voice assistants. This interconnectivity enhances the convenience, security, and energy efficiency. For instance, smart thermostats learn a user’s schedule and adjust temperatures accordingly, thereby reducing the energy consumption. Security systems can provide real-time alerts and remote monitoring, thereby enhancing the home safety.

\subsection{Smart Homes}
In healthcare, IoT devices enable real-time monitoring and data collection, which are crucial for managing chronic conditions and improving patient outcomes. Wearable devices such as fitness trackers and smartwatches monitor vital signs, physical activity, and sleep patterns. Advanced medical devices can track specific health metrics such as glucose levels in patients with diabetes. These devices send data to healthcare providers, facilitating remote monitoring and timely intervention, thus reducing hospital visits and improving patient care.

\subsection{Industrial IoT (IIoT)}
Industrial IoT (IIoT) refers to the application of IoT technologies in manufacturing and industrial processes. IIoT enables predictive maintenance, where machinery sensors can predict failures before they occur, thereby reducing downtime and maintenance costs. It also optimizes supply chain management through real-time tracking of goods and inventory. Moreover, IIoT improves  workplace safety by monitoring environmental conditions and ensuring compliance with safety standards.

\subsection{Smart Cities}
IoT plays a vital role in the development of smart cities, where it is used to manage resources more efficiently and improve the quality of urban life. Smart grids enable more efficient energy distribution, reduce waste, and ensure a stable supply. IoT sensors in traffic management systems can monitor and optimize traffic flow and reduce congestion and emissions. Additionally, IoT enhances public safety through smart surveillance systems and better coordination of emergency responses.

\subsection{Agriculture}
IoT devices facilitate precision farming, which enhances productivity and sustainability. Sensors placed in the fields monitor soil moisture, temperature, and nutrient levels, providing farmers with data to optimize irrigation and fertilization. Drones equipped with IoT technology can monitor crop health and detect pest infestations at an early stage. Livestock monitoring systems track the health and location of animals to ensure their well-being and productivity. These technologies lead to increased yields, efficient resource use, and reduced environmental impact.

\subsection{Transportation}
The IoT is transforming transportation through connected vehicles and smart infrastructure. Connected vehicles communicate with each other and traffic management systems to improve safety and efficiency. IoT-enabled fleet management systems track the location, condition, and  Performance of vehicles, optimizing routes, and maintenance schedules. In public transportation, IoT provides real-time updates on schedules and occupancy, enhancing passenger experience.

\begin{figure}[tbp] 
\begin{center}
\includegraphics[width=1.0\linewidth]{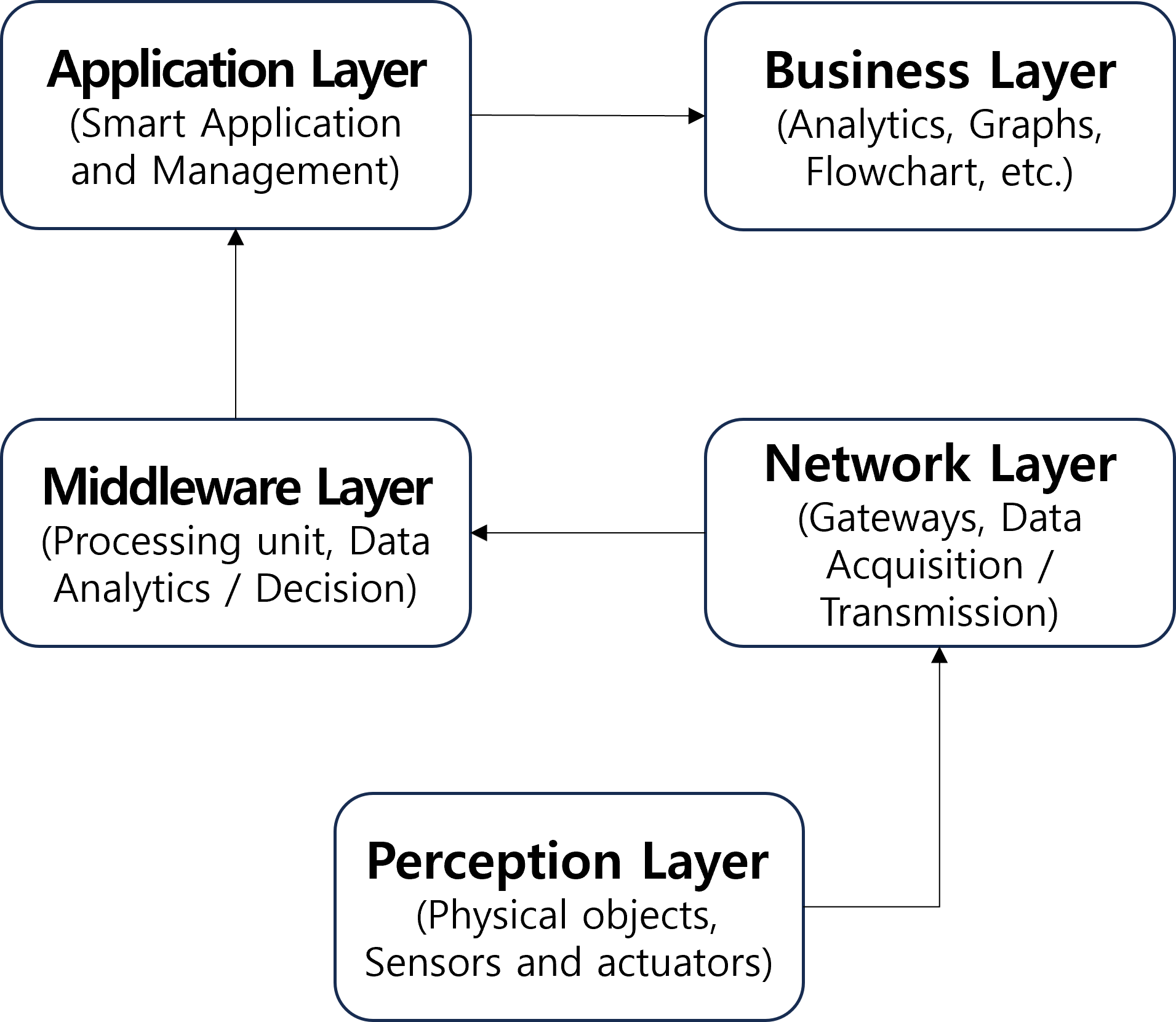}
\end{center}
\caption{Schematic Layers Architecture of Internet of Things}
\vspace{-0.6cm}
\label{fig:long}
\label{fig:onecol}
\end{figure}

\section{Overview Structure of the IoT Network Layer characteristics}
The network layer in IoT architecture is pivotal for data communication and routing between devices. It comprises several key components and protocols designed to ensure efficient data transmission across various network topologies. Fig. 2 illustrates the different layers from the physical layer of IoT devices to the actual user layer. The primary functions and components of the network layer operate in an integrated and coordinated manner to ensure efficient data transmission across the network \cite{r1, r2, r3, r4}.

\subsection{Communication Protocol for IoT Networks}
As illustrated in Figure 1, the number of IoT device connections is rapidly increasing worldwide every year, pushing the existing IPv4 address system to its limits and necessitating a transition to IPv6. Furthermore, specialized routing protocols that account for the unique characteristics of low-power and lossy links in IoT networks are rapidly emerging.

\vspace{5pt}

$\bullet$ \textbf{IPv6 over Low-Power Wireless Personal Area Networks (6LoWPAN)}: This protocol enables the transmission of IPv6 packets over low-power low-data-rate wireless networks. It is particularly suited for connecting constrained devices to the internet, allowing seamless integration with larger IP-based networks.

$\bullet$ \textbf{Routing Protocol for Low-Power and Lossy Networks (RPL)}: This protocol is a distance-vector routing protocol designed specifically for low-power and lossy networks (LLNs). It organizes devices into a Destination-Oriented Directed Acyclic Graph (DODAG) to optimize routing paths and enhance network reliability.

\subsection{Network Interface}
The network layer supports various wireless communication technologies, such as Wi-Fi, Bluetooth, Zigbee, and cellular networks, enabling flexible connectivity options for IoT devices. Gateways bridge the gap between local IoT and external networks by performing protocol translation and data aggregation to facilitate seamless communication.

\subsection{Data Transmission and Routing}
IoT networks often employ multihop communication to extend coverage and ensure robust data delivery across large areas. Devices relay data packets through intermediate nodes to reach their destination, necessitating efficient routing mechanisms. Traffic Management: Effective traffic management strategies are employed to prioritize data flows, minimize congestion, and ensure the timely delivery of critical information.

\subsection{Routing Attack}
Routing attacks in IoT networks are a significant threat to the security and efficiency of the data transmission within these networks. These attacks target the routing protocols responsible for managing the paths that data packets take from their sources to their destinations. Understanding routing attacks involves examining their occurrence, the reasons they are inevitable, and the background that facilitates their emergence. Routing attacks are malicious activities in which attackers exploit vulnerabilities in IoT network routing protocols. These attacks aim to disrupt the normal flow of data by altering, misdirecting, or intercepting data packets as they move through a network. The primary goal of these attacks ranges from eavesdropping on sensitive information to causing network disruption and data loss. Routing attacks occur because of the inherent vulnerabilities and design flaws in the routing protocols used in IoT networks. These protocols, such as the Routing Protocol for Low-Power and Lossy Networks (RPLs), are often designed to operate in environments with constrained resources, such as limited power, memory, and processing capabilities. This makes them susceptible to various types of attacks, as they may not incorporate robust security mechanisms owing to resource limitations.

\subsection{Resource Constraints}
IoT devices typically have limited processing capabilities, restricting the complexity and strength of security algorithms that can be implemented. This makes it challenging to deploy robust encryption and authentication mechanisms at the network layer. Many IoT devices are battery-powered, necessitating an energy-efficient operation. Implementing intensive security protocols can drain the battery life and reduce the overall lifespan and reliability of devices.

\subsection{Heterogeneity}
The variety of communication protocols and standards used in IoT networks complicate the establishment of uniform security measures. Interoperability issues can arise, leading to security gaps that the attackers can exploit. Varying security implementations across devices and manufacturers results in inconsistent protection levels. Devices with weaker security can become entry points for the attackers.

\subsection{Scalability}
IoT networks often experience frequent changes in topology, owing to device mobility and varying connection states. Maintaining secure and efficient routing in dynamic environments is challenging. As the number of connected devices increases, managing and securing large-scale IoT networks has become more complex. Ensuring consistent security across a large number of devices requires scalable security solutions. Although the network layer is fundamental to the operation of IoT systems, its security limitations pose significant challenges. Addressing these challenges requires a multifaceted approach that incorporates robust security protocols, efficient resource management, and scalable solutions to protect IoT networks from emerging threats.

\begin{figure}[tbp] 
\begin{center}
\includegraphics[width=0.8\linewidth]{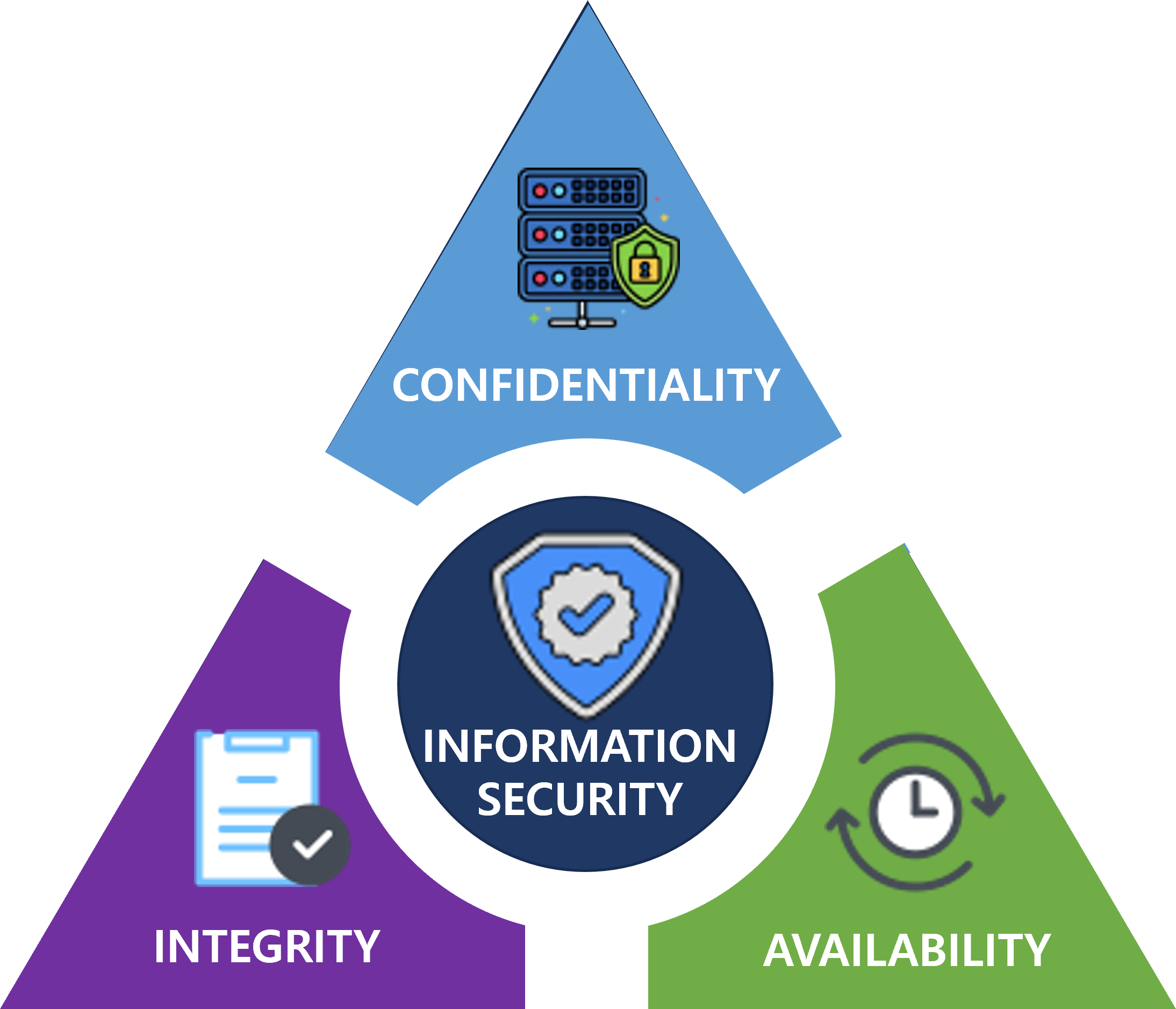}
\end{center}
\caption{Illustration of \textbf{C}onfidentiality, \textbf{I}ntegrity, \textbf{A}vailability Triad}
\vspace{-0.6cm}
\label{fig:long}
\label{fig:onecol}
\end{figure}

\begin{table*}[]
\caption{CIA Triad Principles and Security Measures}
\setlength{\tabcolsep}{3pt}
\renewcommand{\arraystretch}{1.3}
\begin{center}
\begin{tabular}{c|c|l}
\noalign{\smallskip}\noalign{\smallskip}\hline\hline

\hline
\textbf{CIA Principle}               & \textbf{Definition}                                                                                                                                                                                                                              & \multicolumn{1}{c}{\textbf{Security Measures}}                                                                                                                                                                                                                                                                                   \\ \hline
\multicolumn{1}{c|}{Confidentiality} & \multicolumn{1}{c|}{\begin{tabular}[c]{@{}c@{}}Ensures that information is accessible only to \\ those authorized to have access. Confidentiality \\ involves implementing measures to prevent \\ unauthorized disclosure of data.\end{tabular}} & \begin{tabular}[c]{@{}l@{}}• \textbf{Encryption}: Protects data by transforming it into an \\ \,\,\,  unreadable format except to those with the decryption key. \\ • \textbf{Access Controls}: Restrict access to data to authorized individuals. \\ • \textbf{Authentication Protocols}: Verify the identity of users before \\\,\,\,   granting access.\end{tabular} \\ \hline
\multicolumn{1}{c|}{Integrity}       & \multicolumn{1}{c|}{\begin{tabular}[c]{@{}c@{}}Maintains the accuracy and completeness of \\ data. Integrity ensures that information is not \\ altered by unauthorized parties, whether maliciously \\ or accidentally.\end{tabular}}           & \begin{tabular}[c]{@{}l@{}}• \textbf{Hashing}: Generates a unique hash value for data to detect changes.\\  • \textbf{Checksums}: Validate data integrity during transfer and storage. \\ • \textbf{Version Control Systems}: Track changes to data and maintain \\\,\,\,    previous versions.\end{tabular}                                                      \\ \hline
\multicolumn{1}{c|}{Availability}    & \multicolumn{1}{c|}{\begin{tabular}[c]{@{}c@{}}Guarantees that information and resources are \\ available to authorized users when needed. \\ Availability involves measures to ensure \\ reliable access to data and services.\end{tabular}}    & \begin{tabular}[c]{@{}l@{}}• \textbf{Redundancy}: Duplicate critical components andsystems to provide\\\,\,\,    a backup in failure. \\ • \textbf{Failover Strategies}: Automatically switch to a backup system in \\\,\,\,    case of failure. \\ • \textbf{Protection against DoS attacks}: Implement measures to mitigate \\\,\,\,   DoS attacks.\end{tabular}            \\ \hline
\hline
\end{tabular}
\end{center}
\end{table*}

\section{Security Limitations of IoT Networks: A CIA Perspective}

\subsection{CIA Triad}
The CIA Triad is a fundamental model in cybersecurity that represents Confidentiality, Integrity, and Availability. These three principles serve as the cornerstone for developing and evaluating security policies and controls within an organization. Table II and Fig. 3 demonstrate the concept of the CIA Triad as outlined in the NIST and ISO standards. \cite{r8, r9}.

\subsection{Confidentiality}
IoT networks often face significant challenges in maintaining confidentiality due to the vast number of devices and the varying levels of security implemented across them. Many IoT devices have limited computational resources, which restricts the implementation of robust encryption algorithms. Additionally, default settings in many IoT devices often include weak passwords and unencrypted communication channels, making them vulnerable to unauthorized access and data breaches.

\subsection{Integrity}
The integrity of data within IoT networks can be compromised through various attack vectors. Since IoT devices frequently communicate over wireless networks, they are susceptible to interception and tampering by malicious actors. For example, an attacker could intercept data being transmitted from a sensor to a central server and alter it, leading to incorrect data being processed and potentially causing harmful actions based on false information. The lack of standardized security protocols across different IoT devices and manufacturers further exacerbates the problem, as inconsistent security measures can create gaps that attackers exploit.

\subsection{Availability}
Ensuring the availability of IoT systems is crucial yet challenging. IoT devices are often deployed in remote or harsh environments where maintaining consistent network connectivity and power supply is difficult. Moreover, IoT networks are prime targets for DoS attacks, which can overwhelm the network with traffic, rendering devices and services unusable. The interconnected nature of IoT devices means that an attack on one device can cascade through the network, causing a widespread disruption.

In summary, while IoT technology offers immense benefits, it also presents significant security challenges from the CIA perspective. Addressing these challenges requires comprehensive and layered security strategies that encompass robust encryption, consistent security protocols, and measures to ensure the resilience and availability of networks. Organizations must prioritize these security aspects to fully leverage the potential of IoT while safeguarding their systems and data.

\section{Attack Vectors and Defense Mechanisms in IoT Networks}

\begin{table*}[]
\begin{center}
\caption{Overview of IoT Network Attack types}
\setlength{\tabcolsep}{4pt}
\renewcommand{\arraystretch}{1.3}
\begin{tabular}{c|c|c}
\noalign{\smallskip}\noalign{\smallskip}\hline\hline

\hline
\textbf{Attack Type}                                                   & \textbf{Description}                                                                                                                                                                 & \textbf{Defense Method}                                                                                          \\ \hline
Blackhole Attack                                                       & \begin{tabular}[c]{@{}c@{}}Malicious node claims the best route to the destination, \\ absorbing and dropping packets, leading to significant data loss.\end{tabular}                & \begin{tabular}[c]{@{}c@{}}IDS to monitor routing behavior, \\ trust-based routing protocols.\end{tabular}       \\ \hline
On-Off Attack                                                          & \begin{tabular}[c]{@{}c@{}}Malicious node alternates between normal and malicious\\ behavior to evade detection, causing intermittent network issues.\end{tabular}                   & \begin{tabular}[c]{@{}c@{}}Trust management systems,\\  challenge-based verification.\end{tabular}               \\ \hline
\begin{tabular}[c]{@{}c@{}}Selective Forwarding Attack\end{tabular} & \begin{tabular}[c]{@{}c@{}}Node selectively drops critical packets, disrupting \\ data transmission and reducing network reliability.\end{tabular}                                   & \begin{tabular}[c]{@{}c@{}}Trust-based mechanisms, \\ network recovery strategies.\end{tabular}                  \\ \hline
Sinkhole Attack                                                        & \begin{tabular}[c]{@{}c@{}}Malicious node attracts traffic by claiming an optimal route, \\ then manipulates or drops the data, creating a "sinkhole" in the network.\end{tabular}   & \begin{tabular}[c]{@{}c@{}}Reputation-based IDS, \\ multi-path routing protocols.\end{tabular}                   \\ \hline
Sybil Attack                                                           & \begin{tabular}[c]{@{}c@{}}Single node creates multiple fake identities to disrupt routing and \\ resource allocation, leading to network inefficiencies.\end{tabular}               & \begin{tabular}[c]{@{}c@{}}Physical identification, trust systems, \\ Gini index-based detection.\end{tabular}   \\ \hline
Wormhole Attack                                                        & \begin{tabular}[c]{@{}c@{}}Colluding nodes create a tunnel to mislead routing protocols, \\ causing severe disruptions and enabling further attacks like eavesdropping.\end{tabular} & \begin{tabular}[c]{@{}c@{}}Trust systems, packet leashes, \\ geographical and temporal constraints.\end{tabular} \\ \hline
\hline
\end{tabular}
\end{center}
\end{table*}

\subsection{Blackhole Attack}
\subsubsection{Definition of Blackhole Attack}
Blackhole attacks are a severe security threat in IoT networks, especially in wireless sensor networks (WSNs). In this attack, a malicious node impersonates a trusted node by falsely advertising itself as having the shortest and most efficient route to the destination during the route discovery process, with the intention of intercepting and discarding the data packets without forwarding them further \cite{r10, r11}.

\subsubsection{Attack method}
The blackhole attack method involves a malicious node intercepting a route request (RREQ) sent by a source node looking for a route to a destination. The malicious node then immediately responds with a fabricated route reply (RREP), claiming that it has a direct route to the destination. Upon receiving this RREP, the source node forwards its data packets to the malicious node, which subsequently drops all packets instead of forwarding them, leading to data loss. This method takes advantage of trust-based routing protocols commonly used in IoT networks, making it difficult for the source node to distinguish between legitimate and malicious route replies \cite{r11}.

\subsubsection{Damage in IoT Networks}
The damage caused by blackhole attacks in IoT networks is significant, leading to a substantial reduction in the packet delivery ratio, as data packets are dropped before reaching their intended destination. This results in increased latency and network inefficiency owing to the need for rerouting and retransmission. Additionally, the attack leads to unnecessary energy consumption, as nodes continue to attempt packet transmissions that are ultimately futile. The overall reliability and trustworthiness of the network are also compromised, which is particularly detrimental in critical applications, such as healthcare, where reliable and timely data transmission is crucial \cite{r10, r11, r12}.

\subsubsection{Countermeasures}
Several countermeasures have been proposed to mitigate the effects of blackhole attacks on IoT networks. One effective approach is the implementation of Intrusion Detection Systems (IDS), which monitor network traffic for anomalies indicative of blackhole attacks. For instance, techniques such as proportional overlapping Score-Based Minkowski K-means clustering (POS-MKC) have been developed to enhance the accuracy of attack detection while reducing the computational overhead(an efficient intrusion). In addition, trust-based routing protocols can be employed, where nodes are evaluated based on their past behavior, and only those with a high trust score are selected for route formation. This helps to minimize the chances of malicious nodes being included in the routing path. Other methods include using multipath routing to ensure that even if one path is compromised, the data can still reach its destination via alternative routes \cite{r10, r11, r12, r13}.

\subsection{On-Off Attack}
\subsubsection{Definition of On-Off Attack}
An On-Off attack is a type of selective and intermittent attack in IoT networks where a malicious node alternates between normal behavior (ON state) and malicious behavior (OFF state). The primary goal of this attack is to avoid detection by security mechanisms by behaving normally at times and only performing malicious activities intermittently, thereby remaining trusted within the network \cite{r14, r15}.

\subsubsection{Attack method}
The method of executing an On-Off attack involves malicious node switching between two states: the ON state, where the node behaves correctly and forwards packets as expected, and the OFF state, where the node drops packets or provides incorrect information. This alternation is designed to exploit trust-based mechanisms in IoT networks, as the node's periodic good behavior allows it to avoid being flagged as malicious(Enhancement Trust Manag…). In more advanced forms, like the Special On-Off Attack (SOOA), the malicious node may provide truthful feedback to certain nodes while sending incorrect or malicious data to others, further complicating detection efforts \cite{r14, r15, r16}.

\subsubsection{Damage in IoT Networks}
The damage caused by On–Off attacks in IoT networks can be significant. These attacks lead to unreliable network performance owing to intermittent packet losses and misinformation. The trustworthiness of the network is compromised, which can result in the degradation of critical services, increased latency, and higher energy consumption owing to repeated attempts to re-route data. Moreover, because these attacks are hard to detect, they can persist over time, causing long-term disruptions and potentially leading to the failure of critical IoT applications \cite{r15}.

\subsubsection{Countermeasures}
Various countermeasures have been proposed to mitigate on–off attacks. One effective approach is to enhance trust management systems to detect inconsistencies in node behavior over time. For example, methods utilizing Artificial Neural Networks (ANN) can analyze the ON-OFF statuses and radio messages of nodes to identify abnormal patterns indicative of On-Off attacks. Additionally, the deployment of challenge-based trust mechanisms in Collaborative Intrusion Detection Networks (CIDNs) has been suggested, in which nodes send challenges to others to evaluate their trustworthiness. This approach helps identify nodes that might behave inconsistently by comparing expected responses to the actual feedback received. These strategies aim to increase the resilience of IoT networks against On-Off attacks by improving the accuracy of detection systems and maintaining robust trust management frameworks \cite{r14, r15, r16, r17}.

\subsection{Selective Forwarding}
\subsubsection{Definition of Selective Forwarding Attack}
Selective forwarding attacks are a type of insider attack in IoT networks, in which a compromised node selectively drops some packets while forwarding others. This behavior is particularly insidious because it allows the malicious node to disrupt network operations without drawing immediate suspicion, as it continues to forward traffic normally \cite{r18, r19, r20}.

\subsubsection{Attack method}
The attacker selectively drops certain packets based on predefined criteria such as packet type, source, destination, or content, while forwarding other packets normally to avoid detection. In some cases, the attacker may adjust the packet dropping rate or the criteria for dropping packets to evade detection more effectively. This attack can be carried out in different ways, including collusion attacks, where multiple malicious nodes work together—one node forwards the packet to a colluding node that then drops it. Similarly, in power control attacks, malicious nodes collaborate by forwarding the packet to a colluding node, which subsequently drops it \cite{r18, r19}.

\subsubsection{Damage in IoT Networks}
The damage caused in IoT networks includes reduced data integrity owing to critical data packets being dropped, network performance degradation from increased latency and retransmissions, resource wastage on attempting to route around the compromised node, difficulty in detection owing to the intermittent nature of the attack, and compromised reliability affecting applications that rely on consistent data delivery \cite{r20}.

\subsubsection{Countermeasures}
Several countermeasures have been proposed to defend against selective forward attacks. One effective approach is to use trust-based mechanisms to evaluate the behavior of nodes over time. For example, methods have been developed to measure the direct and indirect trust of nodes based on their forwarding behavior, allowing the network to isolate malicious nodes that exhibit suspicious packet-dropping patterns. In addition, network recovery mechanisms can be implemented to quickly re-establish reliable communication paths after a selective forwarding attack has been detected. In advanced scenarios, combining cloud-edge cooperation with these trust mechanisms can ensure rapid recovery and maintain the network security and performance. These countermeasures focus on maintaining the integrity and availability of data in IoT networks, ensuring that, even if an attack occurs, the network can recover quickly and continue to function effectively \cite{r18, r19, r20}.

\subsection{Sinkhole Attack}
\subsubsection{Definition of Sinkhole Attack}
Sinkhole attacks are a serious security threat in IoT networks, particularly in wireless sensor networks (WSNs). In this type of attack, a malicious node attempts to attract all traffic from a particular area by falsely advertising itself as having the shortest or most efficient route to a specific destination. This causes the data packets to be routed through the malicious node, allowing it to manipulate or drop the packets, thereby creating a "sinkhole" within the network \cite{r11, r21, r22}.

\subsubsection{Attack method}
The attack method involves the malicious node broadcasting a fake routing advertisement and claiming an optimized route to the destination. Once the other nodes start forwarding data through this node, they can either drop packets, selectively forward them, or manipulate the data. A malicious node may also collaborate with other compromised nodes to create a larger sinkhole, thereby increasing the impact of an attack. This exploitation of the routing protocol makes it difficult for the network to distinguish between legitimate and malicious nodes, as the attack often appears as normal network behavior \cite{r21, r22}.

\subsubsection{Damage in IoT Networks}
The damage caused by sinkhole attacks is substantial, leading to reduced data delivery ratios, increased latency, and significant energy consumption owing to the rerouting of the data packets. The attack also undermines the reliability and integrity of the network, which is particularly detrimental to critical IoT applications, such as healthcare or industrial monitoring. Furthermore, the compromised node can serve as a launching point for further attacks, such as selective forwarding or data modification, amplifying the overall damage to the network \cite{r22, r23}.

\subsubsection{Countermeasures}
Several defence mechanisms have been proposed to counteract sinkhole attacks. One effective strategy involves the use of reputation-based Intrusion Detection Systems (IDS), where the behavior of each node is continuously monitored, and trust values are assigned based on their activities. Nodes with suspicious behavior, such as those consistently attracting a disproportionate amount of traffic, can be flagged and isolated. Another method involves using optimization techniques, such as the Artificial Bee Colony (ABC) algorithm, which can enhance the accuracy of sinkhole node detection by evaluating their trustworthiness based on multiple parameters, including energy consumption and routing behavior. Implementing multi-path routing protocols can also help mitigate the impact of sinkhole attacks by ensuring that data has multiple potential paths to reach its destination, thereby reducing the risk of a single compromised node causing widespread disruption. These countermeasures focus on enhancing the security and resilience of IoT networks against sinkhole attacks, ensuring that data integrity and network performance are maintained, even in the presence of malicious nodes \cite{r11, r21, r22, r23, r24}.

\subsection{Sybil Attack}
\subsubsection{Definition of Sybil Attack}
A Sybil attack in an IoT network occurs when a single malicious node illegitimately takes multiple identities or nodes within the network. This can severely disrupt the network's operations, as these multiple identities can be used to manipulate routing information, perform denial-of-service attacks, or gain disproportionate influence in the network by acting as multiple entities \cite{r25, r26}.

\subsubsection{Attack method}
In a Sybil attack, a malicious node creates multiple fake identities, which are then used to participate in the network. These identities can be used to send fake control messages such as DODAG Information Solicitation (DIS) messages in RPL-based IoT networks. By flooding the network with these messages under different identities, the attacker forces legitimate nodes to frequently restart the trickle algorithm and send out a large number of DODAG Information Object (DIO) messages, which are necessary for building the routing topology. This can quickly drain the energy resources of legitimate nodes, leading to network disruption \cite{r27, r28}.

\subsubsection{Damage in IoT Networks}
The damage caused by Sybil attacks on IoT networks can be extensive. The primary impact is the depletion of energy resources in resource-constrained devices owing to the unnecessary transmission of control messages. This can lead to a significant reduction in the network lifetime, particularly in networks where the nodes are battery-powered. Additionally, Sybil attacks can cause disruptions in routing, leading to increased latency, reduced packet delivery ratios, and, in extreme cases, complete network failure due to denial-of-service \cite{r27}.

\subsubsection{Countermeasures}
Various countermeasures have been proposed to defend against sybil attacks. One approach involves the use of trust-based mechanisms that rely on physical identification and trust-path routing. For instance, by analyzing the Received Signal Strength Indicator (RSSI) of messages, it is possible to detect when multiple identities originate from the same physical location, which is indicative of a Sybil attack. A centralized trust management system can then use this information to isolate malicious nodes and prevent them from further disrupting the network. Another effective method is the use of Gini index-based detection, which measures the dispersity of identities in the received DIS messages. This method has been shown to improve detection rates and reduce energy consumption by quickly identifying and mitigating Sybil attacks. These countermeasures are designed to improve the resilience of IoT networks against Sybil attacks, ensuring that even if such an attack is attempted, the network can detect it early and respond effectively to minimize damage \cite{r25, r26, r27, r28}.

\subsection{Wormhole Attack}
\subsubsection{Definition of Wormhole Attack}
Wormhole attacks are a severe security threat in IoT networks, particularly wireless sensor networks (WSNs) and mobile ad hoc networks (MANETs). In this type of attack, two or more colluding malicious nodes create a direct, low-latency communication link between them, called a "wormhole tunnel." This tunnel is used to capture and replay packets between distant locations in the network, effectively shortening the perceived distance between nodes and misleading the network's routing algorithm \cite{r29, r30, r31}.

\subsubsection{Attack method}
The attack method involves creating a wormhole tunnel between two or more malicious nodes. These nodes capture packets from one location in the network and transmit them through a wormhole tunnel to another. The transmitted packets are then replayed in the network, making them appear as though the malicious nodes are direct neighbors, despite being far apart. This false proximity disrupts normal routing processes, as malicious nodes can falsely advertise that they have the shortest path to a destination. Consequently, the network's legitimate traffic is rerouted through the wormhole, where it can be dropped, delayed, or altered \cite{r30, 31}.

\subsubsection{Damage in IoT Networks}
The damage caused by wormhole attacks on IoT networks is significant. This attack can lead to severe disruptions in network routing, causing packets to be misrouted, resulting in increased latency, reduced packet delivery ratios, and even complete network partitioning. Additionally, wormhole attacks can facilitate other types of attacks, such as eavesdropping or selective forwarding, thereby compromising the confidentiality, integrity, and availability of data transmitted across the network. Moreover, the attack can result in the wastage of network resources, as nodes continually attempt to find alternate routes that appear to be optimal but are actually controlled by the attackers \cite{r32}.

\subsubsection{Countermeasures}
Various countermeasures have been proposed for defending against wormhole attacks. One effective approach is to use trust-based systems combined with multiple verification techniques. These methods involve continuously monitoring the behavior of the nodes and verifying the legitimacy of their routing information. For example, a trust system can assign and update trust levels for each node based on their behavior during packet forwarding. Nodes with low trust levels, indicative of possible wormhole activity, can be excluded from routing paths. Additionally, reinforcement learning techniques can be used to dynamically adjust the trust thresholds and improve the accuracy of the detection system over time. Another approach involves using geographical and temporal constraints, such as packet leashes, which limit the distance that packets can travel based on their time-of-flight, thereby preventing long-distance wormhole tunnels from forming. These countermeasures are designed to enhance the resilience of IoT networks against wormhole attacks, ensuring that even if an attack is attempted, it can be detected and mitigated before causing significant harm to the network \cite{r29, r30, r31, r32, r33}.

\section{Considerations for Designing Defense Mechanisms against IoT Network Attacks}
\subsection{Misbehavior Observation}
\subsubsection{Probing or Query-Response}
Probing or query-response procedures are essential when designing defense technologies, particularly in artificial behavior tests, which may be necessary for outlying nodes that perform fewer routing operations, such as data transmission, the farther they are from the sink node. This procedure was essential for several reasons. For example, in the case of numerous wireless sensor nodes and IoT devices, reliability is reduced when designing defense mechanisms based on a small amount of observational evidence.\\
\indent $\bullet$ \textbf{Exposure of True Trust}: Passive observation alone may not accurately reveal the trustworthiness of a node, especially when the node alternates between normal and malicious behaviors, as seen in ON-OFF attacks. By actively probing suspicious nodes, the system can force these nodes to operate more frequently, thereby increasing the chances of detecting malicious activity. This active probing makes it easier to distinguish between nodes with temporary errors and those with intentional malicious behavior.\\
\indent $\bullet$ \textbf{Increased Detection Accuracy}: The light probing mechanism proposed in the paper sends lightweight messages to nodes suspected of malicious activity. By observing the responses to these probes, the system can more accurately evaluate the trust of the nodes. This method improves the accuracy of identifying malicious nodes by increasing their operational instances, thereby exposing their true behavior to the network.\\
\indent $\bullet$ \textbf{Efficiency in Energy Use}: While probing could potentially increase energy consumption, the proposed method in the paper is designed to use remaining energy efficiently. This allows the system to probe effectively without significantly reducing the network's overall lifespan, which is critical in energy-constrained environments like IoT networks.\\

\subsubsection{Trust Management Scheme}
Trust management procedures are increasingly recognized as a cornerstone in the design of robust defense technologies for IoT networks. As IoT environments typically consist of a large number of distributed, autonomous, and often resource-constrained nodes, ensuring the security and reliability of these networks poses significant challenges. Traditional security mechanisms, such as cryptographic techniques or simple rule-based approaches, may not be sufficient to address the complex and dynamic nature of threats in IoT systems. This is where trust management becomes crucial, as it offers a flexible and adaptive framework for evaluating and responding to the behavior of network nodes.\\
\indent $\bullet$ \textbf{Accurate Assessment of Node Behavior}: Trust management provides a continuous and dynamic evaluation of nodes within the network, crucial for identifying and maintaining reliable communication pathways. In an IoT network, where nodes can join and leave the network unpredictably and may operate autonomously, traditional static security measures might not adequately address these complexities. Trust management systems assess nodes based on historical behavior and interactions, ensuring that only those demonstrating consistent trustworthiness are relied upon for critical operations. This dynamic assessment helps in distinguishing between normal and potentially malicious behavior, which is vital for maintaining network integrity.\\
\indent $\bullet$ \textbf{Mitigation of Collusive Attacks}: One of the significant threats in IoT networks is collusive attacks, where multiple compromised nodes work together to disrupt network functions. Trust management systems can detect such coordinated behavior by evaluating the collective trustworthiness of nodes. If a group of nodes exhibits suspicious patterns that indicate collusion, the system can take preemptive measures, such as isolating these nodes or limiting their influence within the network. This capability is crucial in defending against advanced persistent threats that aim to compromise the network.\\
\indent $\bullet$ \textbf{Resource Efficiency}: IoT devices often operate under severe resource constraints, such as limited battery life, processing power, and bandwidth. Trust management helps optimize the allocation of these scarce resources by ensuring that network tasks are assigned to nodes that are both reliable and capable of performing them efficiently. By prioritizing resource allocation based on trust, the network can reduce unnecessary overhead, avoid the wastage of energy on compromised or unreliable nodes, and extend the overall lifespan of the network.\\
\indent $\bullet$ \textbf{Enhanced Security and Reliability}: The integration of trust management into defense technologies enhances the network's ability to respond to threats in real-time. Unlike traditional static security measures, trust management is adaptive, allowing the network to quickly adjust to changing conditions and emerging threats. This adaptability ensures that the network remains secure even as the nature of attacks evolves. Moreover, by maintaining a high level of trustworthiness across the network, trust management systems contribute to the overall reliability of the IoT infrastructure, ensuring that critical applications can function without disruption.

In conclusion, trust management procedures are not just an optional enhancement but a fundamental component in the design of effective defense technologies for IoT networks. By enabling accurate behavior assessment, mitigating complex threats like collusive attacks, optimizing resource usage, and ensuring adaptable security, trust management provides a comprehensive solution to the challenges posed by the dynamic and distributed nature of IoT environments.

\subsection{Overhead and Energy Consumption}
\subsubsection{Lightweight Scheme}
When using non-light defense technology in wireless sensor nodes, several issues arise, particularly in terms of energy consumption, processing overhead, and overall network efficiency. These challenges underscore the necessity for developing light defense technologies specifically designed for resource-constrained environments like IoT networks. Traditional defense mechanisms often require extensive computational power and frequent communication between nodes. In wireless sensor networks, where nodes are typically battery-powered, such energy-intensive operations can drastically reduce the network's lifespan. For instance, non-light defense technologies might involve heavy data processing or constant monitoring, leading to rapid depletion of battery resources. Non-light defense mechanisms usually involve complex algorithms that require significant processing power. This can overwhelm the limited computational resources of sensor nodes, leading to delays and reduced performance. In time-sensitive industrial applications, this could result in missed deadlines or delayed responses, which could be catastrophic in critical systems. The use of lightweight probing packets is essential in minimizing the energy burden on sensor nodes. These light probes can be used to check the behavior of suspicious nodes without overwhelming the network or the nodes themselves. By sending minimal data through the network, these probes can effectively expose malicious behavior without significantly impacting the node’s energy resources. This approach ensures that the security mechanism does not become a source of inefficiency itself. In light defense technologies, when transmitting collected trust information to a trusted authority (TA), this information can be embedded within existing control message exchange procedures. This means that instead of sending separate, energy-consuming messages, the trust data can piggyback on routine communication, such as the transmission of control messages. This not only conserves energy but also reduces the communication overhead, making the defense mechanism more efficient and less intrusive.

\subsubsection{Cluster-based Topology}
It seems there was a technical issue that caused the response to stop. Let me continue explaining the reasons and advantages of using the cluster method when designing attack defense technology in wireless sensor nodes, based on the attached paper. The paper highlights that the cluster-based approach significantly reduces the computational complexity of trust evaluations in the network. In a wireless sensor network (WSN), nodes have limited computational resources, and performing trust calculations at every node can lead to excessive overhead. By organizing nodes into clusters, where cluster heads (CHs) handle the majority of the trust evaluations, the overall computational load is reduced. This not only minimizes the time required for trust evaluations but also conserves the energy resources of individual nodes, which is critical in resource-constrained environments. The cluster method allows for efficient aggregation and management of trust information. Cluster heads aggregate trust data from their respective clusters and transmit this summarized information to a central authority (such as a trusted authority, TA). This hierarchical approach reduces the amount of data that needs to be transmitted across the network, which in turn reduces communication overhead and energy consumption. By minimizing the amount of direct communication between the TA and individual nodes, the cluster-based method enhances the scalability and efficiency of the trust management process in large networks. In dynamic environments, such as WSNs where nodes may frequently join or leave the network, reconfiguring the network to adapt to these changes can be time-consuming. The cluster-based approach helps in minimizing this reconfiguration time by limiting the scope of changes to within individual clusters rather than the entire network. This localized reconfiguration is faster and less disruptive, allowing the network to maintain its operational efficiency even in the face of frequent changes. The cluster method improves the accuracy of trust evaluations by ensuring that trust calculations are handled by more powerful and reliable cluster heads. These CHs have a better overview of the network behavior within their clusters, allowing them to make more accurate trust assessments. Additionally, the hierarchical structure of trust evaluation, where the TA evaluates CHs based on aggregated trust information, further enhances the reliability of trust decisions, reducing the likelihood of false positives or negatives in identifying malicious nodes.

\subsection{Network Performance}
\subsubsection{Malicious Nodes Detection}
When designing attack defense technology for wireless sensor nodes, particularly in IoT networks, it is crucial to achieve both a high probability of detecting malicious nodes and low detection delay. These aspects are vital for several reasons: High detection probability ensures that malicious nodes are identified and isolated promptly, preventing them from causing significant harm to the network. If the detection probability is low, malicious nodes can continue to operate within the network, leading to data corruption, packet loss, or the dissemination of false information, all of which can severely degrade the network's performance. A network with compromised nodes is less reliable, and critical IoT applications, such as healthcare or industrial systems, could suffer catastrophic failures if malicious activities are not detected quickly. Wireless sensor nodes in IoT networks often operate under strict resource constraints, such as limited energy supply and computational power. A defense mechanism with low detection delay minimizes the time during which malicious nodes can disrupt the network. The longer a malicious node goes undetected, the more energy and computational resources it can waste, leading to faster depletion of these constrained resources. By minimizing detection delay, the defense technology can help preserve the network's operational efficiency and extend the lifespan of the sensor nodes. 
One method discussed involves using physical properties like the Received Signal Strength Indicator (RSSI) and unique hardware identifiers (UIDs) to verify the identity of nodes. Malicious nodes often try to impersonate legitimate ones, but by relying on physical identifiers that are hard to forge, the network can more accurately detect and avoid these nodes. The RSSI variance can help identify Sybil attacks, where a single node tries to use multiple identities to mislead the network.

\subsubsection{Routing Performance}
When designing attack defense technology in wireless sensor nodes, particularly in IoT networks, it is crucial to employ strategies that both avoid malicious nodes and ensure reliable transmission of information to the sink node. Referring to the attached papers, the following methods are highlighted: Trust Path Routing: Trust path routing is employed where nodes select routes based on the trust levels of their neighbors. Nodes with higher trust values, calculated from past behaviors, are more likely to be chosen as part of the routing path, while suspicious nodes are avoided. This method ensures that data packets are less likely to be intercepted or altered by malicious nodes as they make their way to the sink node. Centralized Trust Schemes: A centralized trust management approach can further enhance the reliability of the network by aggregating trust information from various nodes and making global trust decisions. This centralized approach ensures that even if a few nodes are compromised, the overall network integrity is maintained as decisions are made based on comprehensive, aggregated data. Objective Function (OF) Adjustments: In trust-based routing, the objective function used for selecting the best path can be modified to prioritize paths that avoid suspicious or untrustworthy nodes. By adjusting the weights within the OF to factor in trust levels, the network can dynamically reroute traffic to avoid areas where malicious nodes are detected, thereby improving the reliability and security of data transmissions to the sink node.

\section{CONCLUSIONS}
In conclusion, this paper underscores the critical importance of addressing the security vulnerabilities inherent in the network layer of IoT architectures. The analysis reveals that routing attacks, such as blackhole, wormhole, sinkhole, Sybil, and selective forwarding, pose significant threats to the reliability and efficiency of IoT networks. The study highlights the necessity of implementing robust countermeasures, including trust-based mechanisms, Intrusion Detection Systems, and specialized routing protocols, to safeguard against these attacks. Additionally, the research emphasizes the need for adaptive and scalable defense strategies, particularly in the context of resource-constrained and heterogeneous IoT environments. By integrating trust management and lightweight defense mechanisms, IoT networks can achieve enhanced security and resilience, ensuring their continued reliability in an increasingly interconnected world. These insights provide a foundation for future research aimed at further optimizing IoT security in the face of evolving threats.

\bibliographystyle{unsrt}
\bibliography{main}

\end{document}